\documentclass[%
 reprint,
 amsmath,amssymb,
 aps,
]{revtex4-1}

\usepackage{graphicx}
\usepackage{dcolumn}
\usepackage{bm}

\begin{document}


\title{Three Dimensional Rotations of the Electroweak Interaction}

\author{Mark L. A. Raphaelian}
\affiliation{National Security Technologies, LLC., Livermore CA 94550 }%
 \altaffiliation[Also at ]{https://www.linkedin.com/in/markraphaelian}
 \email{email: raphaeml@nv.doe.gov}


\begin{abstract}
 An extension to the standard electroweak model is presented that is different from previous models. This extension involves a three dimensional rotation of a plane defined by two charge-neutral current axes. In this model the plane also contains the three algebraically coupled SU(2) groups quantization axes that comprise the SU(3) group. This extension differentiates leptonic currents from baryonic currents through hypercolor-charge channels and predicts additional symmetries and currents due to different rotational orientations of the interaction projecting onto the plane defining the SU(3) symmetry group and a U(1) symmetry axis.
\begin{description}
\item[PACS numbers]{02.20.-a, 11.10.Nx, 11.40.-q, 12.10.Dm, 12.60.Cn, 12.90.+b, 14.70.-e, 95.30.Cq}
\end{description}
\end{abstract}

\maketitle

The unification between electromagnetic and weak gauge theories producing the standard electroweak theory occurs via a rotation within a plane defined by two orthogonal gauge fields. These fields are rotated about a perpendicular axis such that their admixtures produce the observed U(1) symmetry (electromagnetic) and SU(2) symmetry (weak) gauge fields. The mixing angle is called the Weinberg angle \cite{mandl1}. Another way to view the above rotation is that of the rotation of a plane containing the quantization axes of the U(1) and SU(2) symmetry groups. In this perspective, the plane containing the two perpendicular axes is rotated about an orthogonal ``line of nodes'' axis in a three dimensional space by the Weinberg angle. The projection of the rotated SU(2) quantization axis onto the non-rotated U(1) quantization axis produces the equivalence between the weak coupling charge and electromagnetic coupling charge.

This work proposes an extension to the standard theory involving a two-dimensional rotation with one angle on a plane to a three-dimensional rotation involving two angles within a three dimensional space. In this rotation, three gauge fields are rotated in space such that admixtures of them produce the observed electromagnetic gauge field and two orthogonal neutral gauge fields. The two neutral gauge fields axes define a plane perpendicular to the U(1) quantization axis and contain the charge quantization axes found in the SU(3) group \cite{Halzen1}. Because this extension contains U(1), SU(2), and SU(3) group structures, this extension has the properties of a unified field theory.

The SU(3) group is defined by the Gell-Mann matrices $\lambda_{i}$ where the subscript runs from 1 to 8. Two of the matrices, $\lambda_{3}$ and $\lambda_{8}$, are diagonal and from them we define the following three ``0-component'' matrices:
\begin{eqnarray}\begin{array}{cl}
\lambda^{12}_{_{0}}\,\,=\,\,\lambda_{3}, & \qquad \lambda^{45}_{_{0}}\,\,=\,\,-\frac{1}{2}\left(+\,\lambda_{3} \,+\,\sqrt{3} \lambda_{8}\right), \\ \\ \mathrm{and} & \qquad \lambda^{67}_{_{0}}\,\,=\,\,-\frac{1}{2}\left(+\,\lambda_{3} \,-\,\sqrt{3} \lambda_{8}\right).
\end{array}
\end{eqnarray}

From these definitions, the three SU(2) groups in a positive cyclic ordering (12:45:67) are:
\begin{eqnarray}\begin{array}{ccc}
(\lambda_{1},\lambda_{2},\lambda^{12}_{_{0}}), &
\,\, (\lambda_{5},\lambda_{4},\lambda^{45}_{_{0}}), &
\,\,\mathrm{and}\quad (\lambda_{6},\lambda_{7},\lambda^{67}_{_{0}}).
\end{array}
\end{eqnarray}
Each of the three groups follows the cyclic intragroup (same axis) SU(2) algebra $[\lambda_{k},\lambda_{l}]=+2i\lambda_{m}\,\,\epsilon_{klm}$ and therefore the three ``0-component'' matrices represent three SU(2) symmetry quantization axes. These three SU(2) symmetry quantization axes lie on the plane defined by the two diagonal SU(3) group matrices $\lambda_{3}$ and $\lambda_{8}$.

In addition to the cyclic intragroup algebra of the above three SU(2) groups, each of these groups couple to one another through an intergroup (different axis) algebraic relationships. These algebraic relationships follow the form of: $[\lambda^{a}_{_{0}},\lambda_{l_{1}}]=\pm i\lambda_{l_{2}}$ where the $l_{1}$ and $l_{2}$ subscripts represent two different non-zero-component members from the same the SU(2) group and $\lambda^{a}_{_{0}}$ is a ``0-component'' member from a different SU(2) group. The importance of intergroup coupling is that each of the SU(2) groups is not an isolated SU(2) group but instead form a SU(3) group structure when coupled in this manner.

Another relationship between the three SU(2) groups is $\lambda\,\times\,\lambda\,\,=\,\,0$ for the zero-component matrices:
\begin{eqnarray}\begin{array}{lll}
\big[\lambda^{12}_{_{0}},\lambda^{45}_{_{0}}\big] = \mathrm{\textbf{0}_{3}}, &
\big[\lambda^{45}_{_{0}},\lambda^{67}_{_{0}}\big] = \mathrm{\textbf{0}_{3}}, &
\big[\lambda^{67}_{_{0}},\lambda^{12}_{_{0}}\big] = \mathrm{\textbf{0}_{3}}.
\end{array}
\end{eqnarray}
The importance of the above commutations is that simultaneous eigenvalues along each of the three SU(2) groups ``0-component'' quantization axes are possible. The eigenvalue measurement along an axis is distinct and independent from the eigenvalue measurements along the different axes. These eigenvalues lie on a two-dimensional plane defined by the three SU(2) quantization axes.

Using the definition $\tau = \lambda/2$, the above groups produce three intrinsic angular momentum (or isospin) groups. Each of these intrinsic angular momentum groups has a ``spin-up'' ($\alpha$) and a ``spin-down'' ($\beta$) basis state defined along one of the quantization axes (12), (45), and (67). In a spherical coordinate system, the three intrinsic angular momentum groups are: ($\tau^{12}_{+},\tau^{12}_{-},\tau^{12}_{_{0}}$), ($\tau^{45}_{+},\tau^{45}_{-},\tau^{45}_{_{0}}$), and ($\tau^{67}_{+},\tau^{67}_{-},\tau^{67}_{_{0}}$) \cite{Pushpa1}. The algebraic properties of these intrinsic angular momentum groups are as follows:
\begin{eqnarray}\begin{array}{rclcrcl}
\left[\tau^{a}_{_{0}}, \tau^{a}_{_{0}}\right] & = & \,\mathrm{\textbf{0}_{3}},
& \,\,\, &
\left[\tau^{a}_{_{0}}, \tau^{b}_{_{0}}\right] & = & \,\mathrm{\textbf{0}_{3}},
\\ \\
\left[\tau^{a}_{_{0}}, \tau^{a}_{_{\pm 1}}\right] & = & \pm\, \tau^{a}_{_{\pm 1}},
& \,\,\, &
\left[\tau^{a}_{_{0}}, \tau^{b}_{_{\pm 1}}\right] & = & \mp\,\frac{1}{2} \tau^{b}_{_{\pm 1}},
\\ \\
\left[\tau^{a}_{_{+ 1}},\tau^{a}_{_{- 1}}\right] & = & -\, \tau^{a}_{_{0}},
& \,\,\, &
\left[\tau^{a}_{_{+ 1}},\tau^{b}_{_{- 1}}\right] & = & \,\mathrm{\textbf{0}_{3}},
\\ \\
\left[\tau^{a}_{_{\pm1}}, \tau^{a}_{_{\pm1}}\right] & = & \,\mathrm{\textbf{0}_{3}},
& &
\left[\tau^{a}_{_{\pm1}}, \tau^{b}_{_{\pm1}}\right] & = & -\,\frac{1}{\sqrt{2}} \tau^{c}_{_{\mp1}},
\end{array}
\end{eqnarray}
where ``$a$,'' ``$b$,'' and ``$c$'' are each one of the three cyclic ordered (12:45:67) SU(2) groups contained within SU(3) and $a\neq b\neq c$ \cite{Raphaelian1}.

As with the standard electroweak theory with its three currents, the above intrinsic angular momentum groups allow us to define eight coordinate style currents of the form $J_{i}^{\mu}(x) \, = \, \overline{\psi}\,\gamma^{\mu}\,\tau_{i}\,\psi$. Of the eight currents, six are associated with the three coplanar SU(2) quantization axes and two with the two SU(3) perpendicular axes defining the plane. Traditionally the six coplanar SU(2) currents are reconfigured into spherical coordinate style currents and coupled to a gauge field to produce the charge-raising and charge-lowering interaction forms:
\begin{eqnarray}\begin{array}{l}
-\,g_{_{p}}\,\sum_{i}\,J^{\mu}_{i}(x)\,G_{i\,\mu}(x) \\ \\
\qquad = \,\, -\,g_{_{p}}\,(\,J^{\mu}_{12}(x)\,G^{\dag}_{12_{\mu}}(x) + J^{\mu^{\dag}}_{12}(x)\,G_{12_{\mu}}(x)\,) \\ \\
\qquad \qquad \,\, -\,g_{_{p}}\,(\,J^{\mu}_{45}(x)\,G^{\dag}_{45_{\mu}}(x) + J^{\mu^{\dag}}_{45}(x)\,G_{45_{\mu}}(x)\,) \\ \\
\qquad \qquad \quad\,\, -\,g_{_{p}}\,(\,J^{\mu}_{67}(x)\,G^{\dag}_{67_{\mu}}(x) + J^{\mu^{\dag}}_{67}(x)\,G_{67_{\mu}}(x)\,),
\end{array}
\end{eqnarray}
where $g_{p}$ is a coupling constant of the interaction in the SU(3) group plane and $G_{a}\,(G^{\dag}_{a})$ is the gauge field associated with the respective current axis. These six interaction terms represent the charge changing currents within the SU(3) symmetry group. These terms also represent the SU(3) interaction current projecting onto a SU(2) symmetry quantization axes.

In addition to the six charge changing current terms, there are two charge-neutral current terms associated with the two orthogonal SU(3) axes defining the plane. These current terms follow the form of the two generators defining the SU(3) group:
\begin{eqnarray}\begin{array}{lcl}
J^{\mu}_{3}(x) & = &
\frac{1}{\sqrt{2}}\,\left(\,\overline{\psi}_{1}\,\gamma^{\mu}\,\psi_{1}-\overline{\psi}_{2}\gamma^{\mu}\,\psi_{2}\,\right), \\ \\
J^{\mu}_{8}(x) & = &
\frac{1}{\sqrt{6}}\,\left( \,\overline{\psi}_{1}\,\gamma^{\mu}\,\psi_{1}+\overline{\psi}_{2}\gamma^{\mu}\,\psi_{2} -2\,\overline{\psi}_{3}\gamma^{\mu}\,\psi_{3} \,\right),
\end{array}
\end{eqnarray}
and couple to fields defined along their respective quantization axis.

As with the charge-changing currents, these two charge-neutral currents can also be written in another more symmetrical manner by defining a charge-raising neutral current and charge-lowering neutral current,
\begin{eqnarray}\begin{array}{c}
J^{\mu^{\pm}}_{38}(x) = \mp\frac{1}{2}(\,J^{\mu}_{3}(x) \,\pm i\,\frac{1}{\sqrt{3}}\,J^{\mu}_{8}(x)\,),
\end{array}
\end{eqnarray}
and an associated charge-raising and charge-lowering gauge field,
\begin{eqnarray}\begin{array}{c}
G^{^{\pm}}_{38_{\mu}}(x) = \mp\frac{1}{\sqrt{2}}\,(\,G_{3_{\mu}}(x) \,\pm i\,G_{8_{\mu}}(x)\,),
\end{array}
\end{eqnarray}
such that,
\begin{align}
& J^{\mu^{+}}_{38}(x)\,G^{^{-}}_{38_{\mu}}(x) \,+\,J^{\mu^{-}}_{38}(x)\,G^{^{+}}_{38_{\mu}}(x) \nonumber \\
& \qquad \qquad \qquad = \,\,
J^{\mu}_{3}(x)\,G_{3_{\mu}}(x)\,+\,J^{\mu}_{8}(x)\,G_{8_{\mu}}(x).
\end{align}
Not only does the equivalence shows that the two neutral currents can also be viewed as a form of charge-raising and charge-lowering currents, but also due to the nature of the SU(3) group structure constants, the current along the one neutral current axis intrinsically has three times the strength of the current along the other axis.

In the proposed extension to the standard electroweak rotation, the total interaction current has a coupling strength $g_{s}$ and projects onto a plane defined by the SU(3) group's neutral current axes and an orthogonal axis to the plane that defines the U(1) quantization axis. Using $g_{_{p}}$ and $g_{_{0}}$ as the coupling constants on the group plane and U(1) axis respectively, and $\alpha$ as the angle that the projection on the plane makes relative to a planar axis, it follows that the interaction current can be written as:
\begin{eqnarray}\begin{array}{l}
-\,g_{_{s}}\,\sum_{i}\,J_{i}^{\mu}(x)\,G_{i\,\mu}(x) \\ \\
\qquad = \,\,  -\,g_{_{p}}\,J^{\mu}_{p}(x)\,G_{p_{\mu}}(x) \,\, -\,g_{_{0}}\,J^{\mu}_{0}(x)\,G_{0_{\mu}}(x), \\ \\
\qquad = \,\,  -\,g_{_{3}}\,J^{\mu}_{3}(x)\,G_{3_{\mu}}(x)
\,\, -\,g_{_{8}}\,J^{\mu}_{8}(x)\,G_{8_{\mu}}(x) \\ \\
\quad \qquad \qquad \qquad \quad \,\, -\,g_{_{0}}\,J^{\mu}_{0}(x)\,G_{0_{\mu}}(x),
\end{array}
\end{eqnarray}
where $g_{_{3}}$ and $g_{_{8}}$ are the coupling constants associated with the projections onto the two SU(3) group's neutral current axes. The coupling constants are defined by the relations:
\begin{eqnarray}\begin{array}{lll}
g^{}_{3} = g_{_{p}}\cos\alpha, & g^{}_{8} = g_{_{p}}\sin\alpha, & \tan\alpha = g^{}_{8}/g^{}_{3}.
\end{array}
\end{eqnarray}

Unlike the standard two-dimensional rotation that allows for a mixture of two gauge fields, the above gauge fields allow for admixtures from three gauge fields through a three-dimensional rotation of the form:
\begin{eqnarray}\begin{array}{rcl}
\left( \begin{array}{ccc} G_{3_{\mu}}(x) \\ G_{8_{\mu}}(x) \\ G_{0_{\mu}}(x)  \end{array} \right)
& = & R(\alpha,\beta,\gamma)\,\left( \begin{array}{ccc} X_{3_{\mu}}(x) \\ X_{8_{\mu}}(x) \\ X_{0_{\mu}}(x) \end{array} \right),
\end{array}
\end{eqnarray}
where $R(\alpha,\beta,\gamma)$ is a standard rotation matrix. The proposed strongomagnetic extension reduces to the standard electroweak rotation in a left handed coordinate system under the rotation matrix $R(Z_{_{1}},Y_{_{2}},Z_{_{3}})=R(0,\beta,0)$.

Following the standard electroweak unification scheme, we define a strongomagnetic hypercharge $X_{sm}$ that couples to the U(1) field. For the extended theory, the same unification methodology results in the relation $Q\,\,=\,\,I_{3}\,+\,X_{sm}$, where $Q$ is the electric charge associated with the U(1) symmetry group, $I_{3}$ is the weak isospin associated with the SU(2) symmetry group, and $X_{sm}$ is the strongomagnetic hypercharge associated with the SU(3) symmetry group. The relation also represents the SU(2) and SU(3) groups' quantum number projections onto a quantization axis having U(1) symmetry. In terms of currents, the relation is equivalent to:
\begin{eqnarray}\begin{array}{c}
J^{\mu}_{_{0}}(x) \,\, = \, J^{\mu}_{X_{sm}}(x) \,\, = \, s^{\mu}(x)/e \,\,-\,I^{\mu}_{3}(x).
\end{array}
\end{eqnarray}

The strongomagnetic hypercharge and the corresponding electric charge and isospin state for first generation fermions is presented in the table~\ref{table1}. It shows that the strongomagnetic hypercharge magnitude is independent of the isospin state of the fermion and depends only on whether the fermion is a baryon or lepton. The polarity of the hypercharge depends on whether the fermion is a particle or anti-particle.
\begin{table}
\caption{This table shows the relationship between the quantum numbers associated with the electric, isospin, and strongomagnetic charges for first generation particles.}\label{table1}
\begin{ruledtabular}
\begin{tabular}{|c|c|c|c|c|}
    &  &  &  & \\
    &  & Electric & Weak & Strongomagnetic \\
    &  & Charge & Isospin & Hypercharge \\
   Particle & Symbol & Q & I$_{_{3}}$ & X$_{_{sm}}$  \\ \hline
    &  &  &  & \\
   anti-down        & $\overline{d}$    & $+1/3$ & $+1/2$ & $-1/6$ \\
   anti-up          & $\overline{u}$    & $-2/3$ & $-1/2$ & $-1/6$ \\
   up               & $u$               & $+2/3$ & $+1/2$ & $+1/6$ \\
   down             & $d$               & $-1/3$ & $-1/2$ & $+1/6$ \\
   &  &  &  & \\
   positron         & $e^{+}$           & $+1$  & $+1/2$ & $+1/2$ \\
   anti-neutrino    & $\overline{\nu}$  & $0$   & $-1/2$ & $+1/2$ \\
   neutrino         & $\nu$             & $0$   & $+1/2$ & $-1/2$ \\
   electron         & $e^{-}$           & $-1$  & $-1/2$ & $-1/2$ \\
\end{tabular}
\end{ruledtabular}
\end{table}

The requirement of not having isospin couple to the hypercharge produces an isospin current comprised from currents along both neutral current axes in the SU(3) group plane. This isospin coupling to the U(1) symmetry axis vanishes at rotational angles $\beta=0,\pm\pi$ and $\alpha=\pm2\pi/3$.
\begin{eqnarray}\begin{array}{l}
g^{}_{_{0}}I^{\mu}_{3}(x) \,=\,g_{_{p}}\,(\cos\alpha\,J^{\mu}_{3}(x)+\sin\alpha\,J^{\mu}_{8}(x))\tan\beta.\quad
\end{array}
\end{eqnarray}
When we use this relation to eliminate the remaining weak isospin current terms that couple to the neutral current axes, $I^{\mu}_{3}(x)\,X_{3_{\mu}}(x)$ and $I^{\mu}_{3}(x)\,X_{8_{\mu}}(x)$, we produce additional crossterms between neutral currents defined along one axis and gauge fields defined along the other axis. To remove these crossterms, the following angular relationship must hold for the coupling coefficients along the two axes defining the plane:
\begin{eqnarray}\begin{array}{lcl}
\tan^{2}\gamma & = & \tan^{2}\alpha , \\
\tan\gamma & = & \pm\,\tan\alpha \,\, = \,\, \pm \, g_{_{8}}/g_{_{3}},
\end{array}
\end{eqnarray}
or $\gamma=\pm\alpha$, from which we can choose the constraints:
\begin{eqnarray}\begin{array}{rclcrcl}
\cos\gamma & = & +\cos\alpha & \quad \textrm{and}\quad \qquad & \sin\gamma & = & \pm\sin\alpha.
\end{array}
\end{eqnarray}

In the electroweak unification, the quantization axis of the SU(2) group is rotated about the ``line of nodes''. The projection of this axis onto the non-rotated U(1) axis allows for the equivalence between the SU(2) coupling constant and the U(1) coupling constant. We can do the same for this rotation by defining the electric charge in terms of the projected coupling coefficients, one from each neutral current axis.
\begin{eqnarray}\begin{array}{l}
g^{}_{_{0}}\cos\beta = g^{}_{_{3}}\cos\gamma\sin\beta + g^{}_{_{8}}\sin\gamma\sin\beta = e
\end{array}
\end{eqnarray}

The above equivalence produces two solutions:
\begin{align}
& (\,g^{}_{_{0}}/e\,) = (\,g^{}_{_{p}}/e\,)\tan\beta \quad \mathrm{for}\,\,\gamma \, = \, +\,\alpha; \nonumber \\
& \qquad \quad= (\,g^{}_{_{p}}/e\,)\tan\beta\cos2\gamma \quad \mathrm{for}\,\, \gamma \, = \, -\,\alpha.
\end{align}

Using the $\gamma =+\alpha$ constraint and identifying $X_{0_{\mu}}(x)$ with $A_{_{\mu}}(x)$, the interaction neutral currents found within the U(1) and SU(3) groups are:
\begin{align}
& -\,g_{_{p}}\,J^{\mu}_{p}(x)\,G_{p_{\mu}}(x) \,\,-\,g^{}_{_{0}}\,J^{\mu}_{X_{0}}(x)\,G_{0_{\mu}}(x) \,=  \nonumber \\
&  \,\, -\,g^{}_{_{p}}/\cos\beta
\left( \begin{array}{l} (\cos^{2}\gamma-\cos\beta\sin^{2}\gamma)J^{\mu}_{3}(x)  \\
\quad -\,(\sin^{2}\beta\cos\gamma)\,s^{\mu}(x)/e\end{array} \right)X_{3_{\mu}}(x)  \nonumber \\
& \,\,\,\,  +\,g^{}_{_{p}}/\cos\beta
\left( \begin{array}{l} (\sin^{2}\gamma-\cos\beta\cos^{2}\gamma)J^{\mu}_{8}(x) \\
\quad -\,(\sin^{2}\beta\sin\gamma)\,s^{\mu}(x)/e\end{array} \right)X_{8_{\mu}}(x) \nonumber \\
& \qquad  -\,s^{\mu}(x)\,A_{_{\mu}}(x).
\end{align}

For the $\alpha= -\gamma$ constraint, the interaction neutral currents found within the U(1) and SU(3) groups are:
\begin{align}
& -\,g_{_{p}}\,J^{\mu}_{p}(x)\,G_{p_{\mu}}(x) \,\,-\,g^{}_{_{0}}\,J^{\mu}_{X_{0}}(x)\,G_{0_{\mu}}(x) \,=  \nonumber \\
&  \,\, -\,g^{}_{_{p}}/\cos\beta
\left( \begin{array}{l} (\cos^{2}\gamma+\cos\beta\sin^{2}\gamma)J^{\mu}_{3}(x)  \\
\,\, -\,(\sin^{2}\beta\cos2\gamma\cos\gamma)\,s^{\mu}(x)/e\end{array} \right)X_{3_{\mu}}(x)  \nonumber \\
& \,\,\,\,  +\,g^{}_{_{p}}/\cos\beta
\left( \begin{array}{l} (\sin^{2}\gamma+\cos\beta\cos^{2}\gamma)J^{\mu}_{8}(x) \\
\,\, -\,(\sin^{2}\beta\cos2\gamma\sin\gamma)\,s^{\mu}(x)/e\end{array} \right)X_{8_{\mu}}(x) \nonumber \\
& \qquad  -\,s^{\mu}(x)\,A_{_{\mu}}(x).
\end{align}

The above SU(3) neutral current terms involve only two angles: $\beta$, the amount of angular rotation about the ``line of nodes'', and $\gamma$, the amount of angular rotation perpendicular to the ``line of nodes.'' Even when we set $\gamma=0$, meaning that the gauge field defined along the $\mathbf{\hat{e}_{8}}$ direction lies on and parallel to the ``line of nodes'' positive directionality, the above total interaction reduces to:
\begin{align}
& -\,g_{_{p}}\,J^{\mu}_{p}(x)\,G_{p_{\mu}}(x) \,\,-\,g^{}_{_{0}}\,J^{\mu}_{X_{0}}(x)\,G_{0_{\mu}}(x) \,\, = \,\, \qquad \qquad \nonumber \\
& \,\, -\,(\,g^{}_{_{p}}/\cos\beta\,)(J^{\mu}_{3}(x)-\sin^{2}\beta\,s^{\mu}(x)/e)\,X_{3_{\mu}}(x) \nonumber \\
& \quad -\,g^{}_{_{p}}\,J^{\mu}_{8}(x)X_{8_{\mu}}(x) \,\, -\,s^{\mu}(x)\,A_{_{\mu}}(x),
\end{align}
and has one more term over what is found in the single neutral current axis case of the standard electroweak theory. In this rotation there is an extra current lying along the ``line of nodes'' that is absent in the standard electroweak theory. This extra current is orthogonal to both the U(1) symmetry and electroweak axes and therefore represents a non-electromagnetic and non-electroweak current. For this rotation, only the gauge field defined along the $\mathbf{\hat{e}_{8}}$ axis does not couple to the electromagnetic current.

At $\gamma=+\pi/2$, the neutral current along the $\mathbf{\hat{e}_{3}}$ direction lies on but anti-parallel to the ``line of nodes'' positive directionality.  Here the above interaction reduces to:
\begin{align}
& -\,g_{_{p}}\,J^{\mu}_{p}(x)\,G_{p_{\mu}}(x) \,\,-\,g^{}_{_{0}}\,J^{\mu}_{X_{0}}(x)\,G_{0_{\mu}}(x) \,\, = \,\, \qquad \qquad \nonumber \\
& \,\, +\,(\,g^{}_{_{p}}/\cos\beta\,)(J^{\mu}_{8}(x)-\sin^{2}\beta\,s^{\mu}(x)/e)\,X_{8_{\mu}}(x) \nonumber \\
& \quad +\,g^{}_{_{p}}\,J^{\mu}_{3}(x)X_{3_{\mu}}(x) \,\, -\,s^{\mu}(x)\,A_{_{\mu}}(x).
\end{align}
As with the $\gamma=0$ rotation, there is an additional term over what is found in the single neutral current axis case and is representative of a non-electromagnetic and non-electroweak neutral current interaction. In this rotation, only the gauge field defined along the $\mathbf{\hat{e}_{3}}$ axis does not couple to the electromagnetic current.

In table~\ref{table1} the strongomagnetic hypercharge magnitude equals $1/2$ for leptons and $1/6$ for baryons. These quantum numbers are equivalent to the normalization constants associated with the generating matrix defining the $J^{\mu}_{3}(x)$ and $J^{\mu}_{8}(x)$ currents. Since these two currents individually couple to the electromagnetic current at interaction rotation angles $\gamma=0,\pm\pi$, and $\gamma=\pm\pi/2$ respectively, then the generators' normalization constants are representative of a scaling factor associated with the interaction currents when projected on the U(1) symmetry axis. The $J^{\mu}_{3}(x)$ interaction current for leptonic interactions and the $J^{\mu}_{8}(x)$ interaction current for baryonic interactions.

Other rotational angles produce interaction current configurations of interest. For $\beta=0$ and $\gamma=0,\pm\pi/2,\pm\pi$, only decoupled neutral-charge SU(3) and electromagnetic interaction currents remain:
\begin{align}
& -\,g_{_{p}}\,J^{\mu}_{p}(x)\,G_{p_{\mu}}(x) \,\,-\,g^{}_{_{0}}\,J^{\mu}_{X_{0}}(x)\,G_{0_{\mu}}(x)\,\,=\,\,\qquad \qquad \nonumber \\
& \qquad \pm\,g^{}_{_{p}}(\,J^{\mu}_{3}(x)X_{3_{\mu}}(x)+J^{\mu}_{8}(x)X_{8_{\mu}}(x)\,)\nonumber \\
& \qquad \quad -\,s^{\mu}(x)\,A_{_{\mu}}(x).
\end{align}
At $\beta=0$ and $\gamma=\pm\pi/4$, the only neutral-charge interaction current that remains is electromagnetic:
\begin{align}
& -\,g_{_{p}}\,J^{\mu}_{p}(x)\,G_{p_{\mu}}(x) \,\,-\,g^{}_{_{0}}\,J^{\mu}_{X_{0}}(x)\,G_{0_{\mu}}(x) \,\, = \,\, \qquad \qquad \nonumber \\
& \qquad \quad -\,s^{\mu}(x)\,A_{_{\mu}}(x).
\end{align}

The full strongomagnetic-electroweak interaction theory has two charge-neutral currents, three charge-raising currents, three charge-lowering currents, and for the constraint $\gamma=+\alpha$ is given by:
\begin{align}
&-\,g_{_{s}}\,\sum_{i}\,J_{i}^{\mu}(x)\,G_{i\,\mu}(x) \,\,= \,\, -\,s^{\mu}(x)A_{\mu}(x) \nonumber \\
&  \,\, -\,g^{}_{_{p}}/\cos\beta
\left( \begin{array}{l} (\cos^{2}\gamma-\cos\beta\sin^{2}\gamma)J^{\mu}_{3}(x)  \\
\quad -\,(\sin^{2}\beta\cos\gamma)\,s^{\mu}(x)/e\end{array} \right)X_{3_{\mu}}(x)  \nonumber \\
& \,\,\,\,  +\,g^{}_{_{p}}/\cos\beta
\left( \begin{array}{l} (\sin^{2}\gamma-\cos\beta\cos^{2}\gamma)J^{\mu}_{8}(x) \\
\quad -\,(\sin^{2}\beta\sin\gamma)\,s^{\mu}(x)/e\end{array} \right)X_{8_{\mu}}(x) \nonumber \\
&\quad \,\, -\,g_{_{p}}\,(\,J^{\mu}_{12}(x)\,G^{\dag}_{12_{\mu}}(x) + J^{\mu^{\dag}}_{12}(x)\,G_{12_{\mu}}(x)\,) \nonumber \\
&\quad \quad -\,g_{_{p}}\,(\,J^{\mu}_{45}(x)\,G^{\dag}_{45_{\mu}}(x) + J^{\mu^{\dag}}_{45}(x)\,G_{45_{\mu}}(x)\,) \,\, \nonumber \\
&\quad \qquad -\,g_{_{p}}\,(\,J^{\mu}_{67}(x)\,G^{\dag}_{67_{\mu}}(x) + J^{\mu^{\dag}}_{67}(x)\,G_{67_{\mu}}(x)\,).
\end{align}

There are many similarities between the interaction current found in the standard electroweak theory and the interaction current found in the strongomagnetic extension to the theory. To achieve the standard electroweak theory limit, $R(0,\beta,0)$, the additional neutral current interaction term must cancel out the additional charge-raising and charge-lowering interaction terms:
\begin{align}
& J^{\mu}_{8}(x)X_{8_{\mu}}(x) \,\, = \nonumber \\
& \,\,\begin{array}{l} (\,J^{\mu}_{45}(x)\,G^{\dag}_{45_{\mu}}(x) + J^{\mu^{\dag}}_{45}(x)\,G_{45_{\mu}}(x)\,) \\
\qquad +(\,J^{\mu}_{67}(x)\,G^{\dag}_{67_{\mu}}(x) + J^{\mu^{\dag}}_{67}(x)\,G_{67_{\mu}}(x)\,).\end{array}
\end{align}

In conclusion we have presented a strongomagnetic extension to the electroweak theory in which a three dimensional rotation is capable of producing a SU(3) group unification with the weak isospin and U(1) groups. In this extension there are three orthogonal neutral current axes, two of which form a two-dimensional plane containing weak isospin, SU(2), and SU(3) algebraic group structures. Unification occurs through neutral current projections onto a U(1) symmetry axis whose isospin independent magnitude, the SU(3) hypercharge, depends on the normalization constant associated with the generator defining the quantization axis. Manifestations of non-electromagnetic and non-electroweak interaction neutral currents occur at specific rotational angles. The extension reduces to the standard electroweak theory in a left-handed coordinate system under a three-dimensional rotation of $R(0,\beta,0)$ and a neutral current interaction term canceling out two associated charge-raising and charge-lowering interaction terms.

\begin{acknowledgments}
This manuscript has been authored by the National Security Technologies, LLC, under Contract No. DE-AC52-06NA25946 with the U.S. Department of Energy. The United States Government and the publisher, by accepting the article for publication, acknowledge that the United States Government retains a nonexclusive, paid-up, irrevocable, world-wide license to publish or reproduce the published form of this manuscript, or allow others to do so, for United States Government purposes. The U.S. Department of Energy will provide public access to these results of federally sponsored research in accordance with the DOE Public Access Plan (http://energy.gov/downloads/doe-public-access-plan), DOE/NV/25946--2648.
\end{acknowledgments}

\begin{thebibliography}\\

\bibitem{mandl1} F. Mandl and G. Shaw, Quantum Field Theory. New York: Wiley, 1986, p271.

\bibitem{Halzen1} F. Halzen and A. D. Martin, Leptons and Quarks, John Wiley \& Sons, New York, NY (1984).

\bibitem{Pushpa1} Pushpa, P. S. Bisht, T. Li, and O. P. S. Negi, Int. J. Theor. Phys., \textbf{51}, 1866 (2012).

\bibitem{Raphaelian1} M. L. Raphaelian, Lecture Notes - Concepts of Rotation Within Quantum Physics, unpublished (2016).

\end{thebibliography}

\end{document}